\documentclass[12pt]{article}

\usepackage{amsmath,amsfonts}
\makeatletter \@addtoreset{equation}{section}

\usepackage{cite}
\usepackage{bm}
\usepackage{dcolumn}
\usepackage[pdftex]{graphicx}
\usepackage{wrapfig}
\usepackage{sidecap}
\usepackage{graphicx}
\usepackage{caption}
\usepackage{subcaption}
\usepackage{graphicx}
\usepackage{dcolumn}
\usepackage{eufrak}
\usepackage{yfonts}
\usepackage{dsfont}
\usepackage{bbold}
\DeclareMathAlphabet{\mathpzc}{OT1}{pzc}{m}{it}
\makeatother
\def\one{{\hbox{ 1\kern-.8mm l}}}
\newcommand{\Dslash}{\not{\hbox{\kern-4pt $D$}}}
\newcommand{\pdslash}{\not{\hbox{\kern-2pt $\partial$}}}
\newcommand{\be}{\begin{equation}}
\newcommand{\bea}{\begin{eqnarray}}
\newcommand{\eea}{\end{eqnarray}}
\newcommand{\ba}{\begin{array}}
\newcommand{\ea}{\end{array}}
\newcommand{\ee}{\end{equation}}

\begin{document}

\begin{titlepage}
\vspace*{1mm}%
\hfill%

\vspace*{15mm}%
\begin{center}

{{\Large {\bf Possible Central Extensions of Non-Relativistic Conformal Algebras in 1+1}}}

\vspace*{15mm} \vspace*{1mm} {Ali Hosseiny }

 \vspace*{1cm}

{\it Department of Physics, Shahid Beheshti University,
G.C., Evin, Tehran 19839, Iran. }

{\it School of Particles and Accelerators, Institute for Research in Fundamental Sciences (IPM)\\
P.O. Box 19395-5531, Tehran, Iran \\}
\vspace*{.4cm}

\vspace*{.4cm}

email: \tt al$\_$hosseiny@sbu.ac.ir

\vspace*{2cm}
\end{center}

\begin{abstract}

We investigate possibility of central extension for non-relativistic conformal algebras in 1+1 dimension. Three different forms of charges can be suggested. A trivial charge for temporal part of the algebra exists for all elements of l-Galilei algebra class. In attempt to find a central extension as of CGA for other elements of the l-Galilei class, possibility for such extension was excluded. For integer and half integer elements of the class we can have an infinite extension of the generalized mass charge for the Virasoro-like extended algebra. For finite algebras a regular charge inspired by Schr\"odinger central extension is possible.

\end{abstract}

\end{titlepage}



\section{Introduction}

Central charge plays a crucial role in connection between 2 dimensional critical phenomena and conformal field theory. The classification of the Virasoro algebra representations is solely done on the basis of the value of c, and the universality classes of critical phenomena in 2d are characterized based on the value of c. In 2d, unitarity and scale invariance is sufficient to ensure conformal invariance; hence conformal invariance of critical phenomena is assured. Wilson uses renormalization group (RG) \cite{Wilson1,Wilson2} to make an encompassing theory of Widom \cite{Widom} and Kadanoff's  scaling hypothesis \cite{Kadanoff}.  Using this frame work he managed to explain many features of critical phenomena. Despite this spectacular success of RG, the spectrum of fixed points was not given; it was conformal field theory (CFT) \cite{Belavin} that succeeded in giving at least a partial answer to this question. 

Local scale invariance, in 2d can be understood as invariance under any (Anti) holomorphic transformation on the complex plane. The generators of this symmetry form the Witt algebra:
\bea\begin{split}
&[L_m,L_n]=(m-n)L_{m+n}.
\end{split}\eea
Other commutators are similar for the chiral elements and cross commutators are trivial. However the quantum theory gives rise to an anomaly for the operator product expansion of the energy-momentum tensor, leading to a central charge in the above algebra. Hence the Virasoro algebra appears :
\bea\begin{split}\label{cftc}
&[L_m,L_n]=(m-n)L_{m+n}+\frac{C_R}{12}n(n^2-1)\delta_{n,-m}
\end{split}\eea 
with a similar change in the chiral component and a separate central charge  $C_L$ . Imposing unitarity requires the two central charges to be equal. Interestingly, requiring the representations of the Virasoro algebra representations to be finite dimensional leads to the minimal series. This is a series of CFTs which have  :
\bea
C=1-\frac{6}{m(m+1)} ,\;\;\;\;\;\;\;\;\;\;\;\;\;\;\;m=2,3,... \;\;\;\;\;\;
\eea
In this way a whole spectrum of critical points appear \cite{Difrancesco}. For example m=3 or c=1/2 corresponds to the critical point of the Ising model \cite{Henkelbook}. The obvious question is can one do the same for non-relativistic conformal algebras. Interest in non-relativistic conformal algebras has recently grown. They appear as the asymptotic symmetry in AdS/CFT construction \cite{Son}-\cite{Barnich2}, mathematicians have paid attention to the mathematical structure of these algebras (See for example \cite{Fushchych} and references there in), and in condensed matter physics many applications have been suggested \cite{Henkelp,Henkellog}.

The best known non-relativistic conformal algebra is the Schr\"odinger algebra which describes symmetries of the Schr\"odinger equation \cite{Hagen,Niederer}. Beside Galilean invariance this algebra respects scaling invariance:
\bea
x\rightarrow\lambda x,\;\;\;\;\;\;\;\;\;\;\;\;\;\;\;\;t\rightarrow\lambda^z t.
\eea
Unlike the standard CFT, here we have anisotropic scaling in space and time with an index $z$. Another conformal algebra in is the Conformal Galilean Algebra (CGA) which in $2D$ is obtained by a contraction of the conformal algebra \cite{Havas}. 

Both Schr\"odinger algebra and CGA belong to a class of non-relativistic algebras named l-Galilei algebras \cite{Henkelprl,Negro}. These algebras fuse Galilean invariance, causality and scaling in space and time. Each member of this class is characterized by an index "l" related to the dynamic exponent via z=2/l. Among these algebras Schr\"odinger algebra and CGA are most explored \cite{Duval1990}-\cite{Aizawa}. In Schr\"odinger algebra, we have an infinite Virasoro like extension which is called Schr\"odinger-Virasoro algebra \cite{Henkel1994}. This algebra has a "Mass" central extension which helps to realize the representation theory of the algebra. Besides, it is conserved as a superselection rule in correlation functions.
For CGA the story of central charge is interesting. In 1+1 dimensions a satisfactory
understanding of this algebra exists \cite{Bagchi2d}. In 2+1 dimensions, there is a known central charge which is called "Exotic" \cite{Lukierski1}. Physical realization of this charge and symmetry has been of interest. The other elements of l-Galilei algebras are less known\footnote{For a discussion on physical significance of elements with $l=4$ and $l=6$ see \cite{{Pleiming},{Henkelpheno}}}. Even their representation theory and correlation functions have not yet been worked out. Concerning central charge some progress has been made
by Martelli and Tachikawa \cite{Tachikawa}. 

In this paper we try to find possible central extensions for these symmetries.
This paper is organized as follows. In section 2 we review Schr\"odinger algebra and its central extension. We then look at CGA in section 3. The core of our paper is in section 4 where we have investigated possible central extension on all element of l-Galilei algebra in 1+1 dimension. We
conclude our paper in section 5.


\section{Schr\"odinger Algebra and Mass Central Charge}

Schr\"odinger algebra represents symmetries of Schr\"odinger equation. Similar to all other non-relativistic conformal algebras, generators of Galilean transformation:
\bea
P_i=-\partial_i,\;\;\;\;\;\;\;\;\;\; H=-\partial_t,\;\;\;\;\;\;\;\;\;\;B=-t\partial_i,\;\;\;\;\;\;\;\;\;\;J_{ij}=-(x_i\partial_j-x_j\partial_i),
\eea
are included in this algebra. Beside these generators, Schr\"odinger algebra consists two other generators which are dilation
\bea
D=-(t\partial_t+\frac{1}{2}x_i\partial_i),
\eea
and Special Schr\"odinger Transformation (SCT):
\bea
K=-(tx_i\partial_i+t^2\partial_t).
\eea
These generators produce transformations as:
\bea
\vec x\rightarrow\frac{{\cal{R}}\vec x+\vec v t^2+\vec b}{\gamma t+\delta},\;\;\;\;\;\;\;\;\;\;\;\;t\rightarrow \frac{\alpha t+\beta}{\gamma t + \delta}
\eea
in which ${\cal{R}}\in SO(D)$ and $\alpha \delta - \beta \gamma=1$. 
It is known that the algebra in its current form is still incomplete to thoroughly represent symmetries of Schr\"odinger equation. If one considers a plane wave $\psi_{E,\vec p}=a\exp(iEt-i\vec p.\vec x)$, the action of translation on $\psi$ is defined as:
\bea
T_{P(\vec b)}\psi(\vec x)=\psi(P(-\vec b)(\vec x,t)).
\eea
It is observed that as one expects the result is another solution of Schr\"odinger equation. For the action of boost however a problem has been arisen. For a boost $\vec v$ it is expected that
\bea
T_{B(\vec v)}\psi(\vec x)=\psi(B(-\vec v)(\vec x,t)).
\eea
The new $\psi$ however is not a solution of Schr\"odinger equation. To be a solution the action of boost on $\psi$ is redefined as:
\bea\label{trans}
T_{B(\vec b)}\psi(\vec x)=\exp(-im\vec v .\vec x+\frac{i}{2}m\vec v^2t)\psi(B(-\vec v)(\vec x,t)).
\eea 
The new $\psi$ is another solution of Schr\"odinger equation and as it is expected, it represents a plane wave with new energy and momentum in boosted coordinate set:
\bea
\vec P \rightarrow \vec P+m\vec v,\;\;\;\;\;\;\;\;\;\;E\rightarrow \frac{{(\vec P+m\vec v)^2}}{2m}.
\eea
Now a new problem arises and according to transformation (\ref{trans}), $T_{P(b)}$ and $T_{B(\vec v)}$ do not commute:
\bea
T_{P(b)}T_{B(\vec v)}=T_{P(b)}T_{B(\vec v)}\exp(im\vec v .\vec b).
\eea
So, to realize a proper transformation of solutions of Schr\"odinger equation it has been found that boost and translation in space should not commute and thereby central extension of Schr\"odinger algebra emerges:
\bea\label{mass}
[B_i,P_j]=M\delta_{ij}.
\eea
 To realize this central charge as a generator a new coordinate $\xi$ is borrowed and a new field $\phi$ is defined which is a Fourier transform of $\psi$ with respect to mass \cite{UnterbergerHenkel}:
\bea
\psi(\vec x,t)=\frac{1}{\sqrt {2\pi}} \int_R d\xi e^{-im\xi}\phi(\xi,\vec x,t)
\eea
Now, Schr\"odinger equation for $\psi$ in $\mathds{R}^{d,1}$ space turns to Klein-Gordon equation for $\phi$ in $\mathds{R}^{d+1,1}$ space:
\bea
2\frac{\partial^2\phi}{\partial t\partial\xi}+\nabla^2\phi=0.
\eea
Metric for this space is:
\bea
ds^2=2dtd\xi+\vec {dx}.\vec{dx}
\eea
Under boost transformation $\xi$ now transforms as:
\bea
\xi\rightarrow\xi+\vec v.\vec x-\frac{1}{2}\vec v.\vec v t.
\eea
The action of algebra on $\phi(\xi,x,t)$ now reads as:
\bea\begin{split}
&B=-t\partial_x+x\partial_\xi,\;\;\;\;\;\;\;\;\;\;\;\;\;\;\;\;\;\;\;\;\;\;\;\;\;\;\;\;\;\;\;M=\partial_{\xi},\cr&D=-2t\partial t-x\partial_x+\Delta,\;\;\;\;\;\;\;\;\;\;\;\;\;\;\;\;\;\;\;\;\
K=-2t^2\partial_t-2tx\partial_x+x^2\partial_\xi-2\Delta t.
\end{split}\eea
in which $\Delta$ is scaling weight. The set of operators H, D and K form an $SL(2,R)$ algebra. So, the algebra can be extended in this direction to form Schr\"odinger-Virasoro algebra \cite{Henkel1994}. The central charge then needs to be extended as well and then we have a new set of operators for M. The infinite extended algebra then can be realized as:
 \bea\begin{split}\label{schrodigerinfinite}
&T^n=-t^{n+1}\partial_t-\frac{(n+1)}{2}t^nx\partial_x-\frac{(n+1)}{2}\Delta t^n+\frac{n(n+1)}{4}t^{n-1}x^2\partial_\xi \cr&P^q=-t^{q+1}\partial_x+(q+1)t^qx\partial_\xi\cr&M^n=t^n\partial_{\xi}
\end{split}\eea
in which $n\in \mathds{Z}$ and $q\in \mathds{Z}+\frac{1}{2}$. The non-vanishing commutators of these generators are:
 \bea\begin{split}\label{schrodingermass}
&[T^m,T^n]=(m-n)T^{m+n}\;\;\;\;\;\;\;\;\;\;\;\;\;\;\;\;\;[T^m,P^q]=(\frac{m}{2}-q)P^{m+q} \cr&[T^m,M^n]=-nM^{m+n},\;\;\;\;\;\;\;\;\;\;\;\;\;\;\;\;\;\;\;\;[P^p,P^q]=(p-q)M^{p+q}
\end{split}\eea
Now consider a scaling field with a weight $"h"$. Borrowing state-operator correspondence from CFT for scaling fields we expect
\bea
T^0|h\rangle=h|h\rangle
\eea
Now, Eq. (\ref{schrodingermass}) suggests other operators play the ladder role. For a conclusive study on representations, Kac determinants etc. see \cite{CRoger,RogerUnterberger}.


\section{Conformal Galilean Algebra}

Conformal Galilean algebra (CGA) was found in \cite{Havas} and then was rediscovered in \cite{Henkelprl} and independently in \cite{Negro}. In $d$ dimension this algebra can be obtained from contraction of conformal algebra. In contraction we let:
\bea\label{contraction}
x\rightarrow x/c\;\;\;\;\;\;\;\;\;\;\;\;\;\;\;\;\;t\rightarrow t\;\;\;\;\;\;\;\;\;\;\;\;\;\;\;\;\;c\rightarrow \infty
\eea
Under such contraction, Poincare transformations reduces to Galilean transformation and conformal symmetry reduces to CGA. Physically we expect it means investigating a system under conformal symmetry in low speed or energy. The algebra consists of:
\bea \label{cga} \begin{split}
&P_i=-\partial_{x_i}\;,\;\;\;\;\;\;\;\;\;\;\;\;\;\;\;\;\;\;\;\;K_i=-t\partial_{x_i},\;\;\;\;\;\;\;\;\;\;\;\;\;\;\;\;\;\;\;\;\;\;\;\;\;\;\;F_i=-t^2\partial_{x_i},   
\cr& H =-\partial_t\;,\;\;\;\;\;\;\;\;\;\;\;\;\;\;\;\;\;\;\;\;\;D=-(t\partial_t+x\partial_x),\;\;\;\;\;\;\;\;\;\;\;\;\;\;\;\;\;\; C=-(2tx\partial_x+t^2\partial_t),
\cr&J_{ij}=-x_i\partial_{x_j}+x_j\partial_{x_i}.
\end{split}\eea 
In 2+1 dimensions the algebra admits a central charge $\Xi$\cite{Lukierski1}
\bea
[K_i,K_j]=\Xi\epsilon_{ij},\;\;\;\;\;\;\;\;\;\;\;\;\;\;\;\;[P_i,F_j]=-2\Xi\epsilon_{ij} \;\;\;;\;\;\; i,j=1,2,
\eea
which is called exotic in the literature. The physical significance of this charge and its representation theory has been of interest\cite{DuvalHorvathy}-\cite{henkelhosseiny}. 

Dilations operator "D" and two other operators (C and H) obey SL(2,R) commutation relations. The algebra can be extended in this direction and similar to the Schr\"odinger one we have a Virasoro like extension of CGA which is called full CGA in the literature: 
\bea\label{gcageneratrs}\begin{split}
&T^n=-t^{n+1}\partial_t-(n+1)t^nx^i \partial_{x_i},
\cr& P^n_i=-t^{n+1}\partial_{x_i}\cr&J_{ij}=-x_i\partial_{x_j}+x_j\partial_{x_i}.
\end{split}\eea
In 1+1 we can obtain this algebra from contracting $CFT_2$ generators\cite{Hosseiny}. Consider elements of Virasoro algebra in complex plane (\ref{cftc}). Now, under contraction (\ref{contraction}) and redefining operators as:
\bea\label{contractoperator}\begin{split}
&T^n=L^n+\overline{L}^n
\cr& P^n=\frac{1}{c}(L^n-\overline{L}^n)
\end{split}\eea
we end up with full CGA in (\ref{gcageneratrs}). Now, consider central charges of $CFT_2$ in (\ref{cftc}). Under contraction limit we are left with new charges for full CGA \cite{{CRoger2},{Henkelp}}:
\bea\label{contractoperator}\begin{split}
&[T^m,T^n]=(m-n)T^{m+n}+\frac{1}{12}C_Tm(m^2-1)\delta_{m,-n},
\cr& [P^m,P^n]=0,
\cr&[T^m,P^n]=(m-n)T^{m+n}+\frac{1}{12}C_Pm(m^2-1)\delta_{m,-n},
\end{split}\eea
Through this contraction $L_0$ and $\bar L_0$ go to $T^0$ and $P^0$. Now if we suppose that as the generators, representations of full CGA can be obtained from contracting the representations of $CFT_2$ too, then we are done with quantizing CGA. Generally when an algebra is contracted there is no necessity that its representations as well can be contracted to yield representations of the obtained algebra. For CGA however, this is the case. To find justifications in this regard see \cite{Bagchi2d}. Under such procedure, representations with charges as $C_T$ and $C_P$ in CGA are obtained from representations of CFT with charges equal to:
\bea\label{}\begin{split}
&C_T=\frac{C_R+C_L}{2},
\cr& C_P=\frac{C_L-C_R}{2c}.
\end{split}\eea
All other interesting elements such as null vector, Kac determinant etc. can be inherited from $CFT_2$ through contraction \cite{Bagchi2d}.


\section{The general class of $l\_ Galilei$ Algebra }

Schr\"odinger algebra and CGA belong to a general class of non-relativistic algebras named $l\_$Galilei algebra. This algebra has been derived from a phenomenological approach by Henkel\cite{Henkelprl} and from algebraical view by Negro et al. \cite{Negro}. If we look for an algebra which let global conformal transformation of time, an scale invariance in space and as well phenomenological considerations we end up with an algebra produced by generators:
\bea \label{lgalilei} \begin{split}
&H =-\partial_t\;,\;\;\;\;\;\;\;\;\;\;\;\;\;\;\;\;\;\;\;\;\;\;\;C=-(2ltx\partial_x+t^2\partial_t),\;\;\;\;\;\;\;\;\;\;\;\;\;\;\;\;\;D=-(t\partial_t+lx\partial_x)\cr&
P^q_i=(-t)^q\partial_{x_i},\;\;\;\;\;\;\;\;\;\;\;\;\;\;\;J_{ij}=-x_i\partial_{x_j}+x_j\partial_{x_i}.
\end{split}\eea 
where $l\in\frac{\mathbb{N}}{2}$ and $q=0,...,2l$. 
These operators produce transformations as:
\bea \label{} \begin{split}
&\vec x\rightarrow\frac{{\cal R}\vec x+t^{2l}\vec c_{2l}+...+t\vec c_1+\vec c_0}{(\gamma t+\delta)^{2l}}\;\;\;\;\;\;\;\;\;\;\;\;\;\;\;\;\;\;\;\;t\rightarrow\frac{\alpha t+\beta}{\gamma t + \delta}
\end{split}\eea 
where ${\cal R} \in SO(d), \vec c_n\in\mathbb{R}$ and $\alpha\delta-\beta\gamma=1$. Dilation operator produces an anisotropic rescaling of space and time:
\bea
\vec x\rightarrow \lambda x\;\;\;\;\;\;\;\;\;\;\;\;\;\;\;\;\;\;\;\;\;\;\;\;\;\;\;\;\;\;\;\;\;\;t\rightarrow\lambda^zt,
\eea
where $z=\frac{1}{l}$. Schr\"odinger algebra is the first element of this class or $l=\frac{1}{2}$. The second element $l=1$ is CGA. Other elements are not well-known as their former ones. Non-zero commutators of the algebra reads as:
\bea \label{lgalilei} \begin{split}
&[D,H]=H,\;\;\;\;\;\;\;\;\;\;\;\;\;\;\;\;\;\;\;\;\;\;\;\;\;[D,C]=-C,\;\;\;\;\;\;\;\;\;\;\;\;\;\;\;\;\;\;\;\;\;\;\;[C,H]=2D,\cr&[J_{ij},P^q_k]=\delta_{ik}P^q_j-\delta_{jk}P^q_i\;\;\;\;\;\;\;[H,P^q_i]=-qP^{q-1}_i,\;\;\;\;\;\;\;\;\;\;\;\;\;\;\;[D,P^q_i]=(l-q)P^q_i,\cr&[C,P^q_i]=(2l-q)P^{q+1}.\;\;\;\;\;\;\;\;\;[J_{ij},J_{kl}]=\delta_{ik}J_{jl}+\delta_{jl}J_{ik}-\delta_{il}J_{jk}-\delta_{jk}J_{il}.
\end{split}\eea 

In all elements of this class of symmetries still $H,D$ and $C$ form an $SL(2,R)$ algebra. So, we can have a loop-algebra-like infinite extension in this direction:
\bea \label{virasorogalilei} \begin{split}
&T^n=-t^{n+1}\partial t-l(n+1)t^nx_i\partial_i,
\cr&P^q_i=-t^{q+l}\partial_i,
\cr&J^n_{ij}=-t^n(x_i\partial_j-x_j\partial_i)
\end{split}\eea 
in which $n\in\mathbb Z$ and $q\in\mathbb Z+l$. Since $P^q$ act basically on space we have called them {\it{spatial operators}} in this note. To distinguish $T^n$ generators we have called them $temporal$ ones. These generators produce the following algebra:
\bea \label{} \begin{split}
&[T^m,T^n]=(m-n)T^{m+n},\;\;\;\;\;\;\;\;\;\;\;\;\;\;\;\;[T^m,J^n_{ij}]=-nJ^n_{ij},
\cr&[T^m,P^q_i]=(lm-q)P^{m+q}_i,\;\;\;\;\;\;\;\;\;\;\;\;\;\;\;[J^m_{ij},P^q_k]=\delta_{ik}P^{m+q}_j-\delta_{jk}P^{m+q}_i,
\cr&[J^m_{ij},J^n_{kl}]=\delta_{ik}J^{m+n}_{jl}+\delta_{jl}J^{m+n}_{ik}-\delta_{il}J^{m+n}_{jk}-\delta_{jk}J^{m+n}_{il},
\end{split}\eea 

Note that the finite subalgebra (\ref{lgalilei}) can be reached via elements with $m=0,\pm 1$ and $q=-l.-l+1,...,l-1,l$ in (\ref{virasorogalilei}).

 Investigating central extension for these algebras, it has been suggested that only two forms of central extension can exist \cite{Tachikawa}. For any $d$ and half integer $l$ we can have a charge as
\bea
[P^p_i,P^q_j]=I^{pq}\delta_{ij}M,
\eea
in which $I^{pq}$ is an antisymmetric tensor. This is an extension of the mass charge for all half integer elements of l-Galilei algebra. For $d=2$ and integer $l$ we can have an extension of Exotic charge:
\bea
[P^p_i,P^q_i]=I^{pq}\epsilon_{ij}\Xi.
\eea
where $I^{pq}$ is symmetric now. Though these charges have been suggested as the most general case for all elements of l-Galilei algebra, if we look at the case of CGA we find that other forms of the central charges can exist in 1+1. Now, we follow on to check if we can find different charges for theses algebras in 1+1. This dimension is unique since we have no rotation and theorem presented in \cite{Tachikawa} does not apply. 

Concerning the central charge basically we can have three forms of central extension:\\
 \\
1:An extension in which the charge is outcome of commutators of temporal operators which we call T charge. \\
\\
2:Another extension that let the charge results from commutating temporal and spatial operators which we have called a B charge in this note.\\
\\
3:A charge that can be an outcome of commutators of spatial operators. Since Schr\"odinger algebra's mass central charge belongs to this category, we have called them M charge.\\

The T charge trivially exist for any element of the l-Galilei class. It comes from the fact that for these algebras the temporal generators always form a $SL(2,R)$ subalgebra. So, we have:
\bea\label{tcharge}
[T^m,T^n]=(m-n)T^{m+n}+\frac{1}{12}C_Tm(m^2-1)\delta_{m+n,0}.
\eea
Now, let's follow on to investigate if a charge can be obtained through commutation of $T^m$ and $P^q$ for arbitrary $l$. We have two forms of $l$: half integer $l$ and integer $l$.\\ \\

\underline{\bf Case 1a}: Integer $l$ for B central charge\\
\\
We should consider that we have such charge for CGA (\ref{contractoperator}) which belongs to $l=1$ of l-Galilei class. Now, one may wonder if such charge can be defined for other elements of the class. We desire the central charge obeys the equation:
\bea
[T^m,P^n]=(lm-n)P^{m+n}+C_Bf(m)\delta_{m+n,0}.
\eea
Now, let's write Jacobi identity:
\bea
[[T^m,T^n],P^q]+[[T^n,P^q],T^m]+[[P^q,T^m],T^n]=0.
\eea
It leads to the following equation for the central charge:
\bea
[(m-n)f(m+n)+(q-ln)f(m)+(lm-q)f(n)]C_B\delta_{m+n+q,0}=0,
\eea
and thereby:
\bea
(m-n)f(m+n)-((l+1)n+m)f(m)+((l+1)m+n)f(n)=0.
\eea
Now, one can easily check that if we suppose $f$ is a polynomial function then we will have two choices. For $l=1$ we have CGA choice or
\bea
f(m)=am+bm^3.
\eea
For other integer $l$ we only have a choice as $f(m)=am$.
So, for $l>1$ we are left with the following possibility:
\bea\label{bcharge}
[T^m,P^q]=(lm-q)P^{m+q}+C_Bm\delta_{m+q,0}.
\eea
At first it seems that we have figured a new charge, but actually this charge can be absorbed in $P^0$ via redefinition $P^0+\frac{C_B}{l+1}\rightarrow P^0$. \\

\underline{\bf Case 1b}: Half integer $l$ for B central charge\\
\\

While temporal operators are indexed by integer numbers, spatial operators are indexed by half integer numbers. So, one may seek a central charge as 
\bea\label{case2b}
[T^m,P^q]=(lm-n)P^{m+q}+C_Bf(m)\delta_{m+2zq,0},
\eea
in which $z$ is an integer number. Now, let's see if Jacobi identity are satisfied by this equation. We desire 
\bea
[[T^m,T^n],P^q]+[[T^n,P^q],T^m]+[[P^q,T^m]T^n]=0,
\eea 
which reduces to:
\bea
(m-n)f(m+n)\delta_{(m+n+2qz),0}+(q-ln)f(m)\delta_{m+2zn+2zq,0}+(lm-q)f(m+q)\delta_{n+2zq+2zm,0}=0.
\eea
Now, we choose $m=-2qz-n$ and thereby we have:
\bea\begin{split}
&(2qz-2n)f(-2qz)+(q-ln)f(-2qz-n)\delta_{n(2z-1),0}\cr&\;\;\;\;\;\;\;\;\;\;\;\;\;\;\;\;\;\;\;\;\;\;\;\;\;\;\;\;\;\;-(l(2qz+n)-q)f(-2qz-n+q)\delta_{n+2z(q-2qz-n),0}=0.
\end{split}\eea
This equation should hold for any "n". So, we have no choice but $f(-2qz)=0$. Thereby any form of central charge as equation (\ref{case2b}) is impossible. 

As it can be seen the supposed to be B charge only exist for $l=1$. Now, we follow to investigate if the algebras admit central extension for commutators of spatial operators. Inspired by Schr\"odinger $mass$ central extension we have called such extension as M charge in this note.\\

\underline{\bf Case 2a}: M Charge for infinite-extended l-Galilei algebra\\

Schr\"odinger algebra has a M charge for commutators of its spatial operators $B$ and $P$ (\ref{mass}). In the compact form of equation (\ref{virasorogalilei}) for this charge we have:
\bea
[P^{\frac{1}{2}},P^{-\frac{1}{2}}]=M.
\eea
One may expect an extension as 
\bea\label{meremass}
[P^q,P^r]=Mf(q)\delta_{q+r,0}
\eea
for infinite extended algebra. The Schr\"odinger-Virasoro however has a different form of M charge in (\ref{schrodingermass}). Instead of having only one operator to embed mass charge to infinite extension of Schr\"odinger algebra we need to define an infinite set of operators (\ref{schrodigerinfinite}). To observe that a central extension as of (\ref{schrodigerinfinite}) can't hold for any element of the l-Galilei class we write Jacobi identity for it:
\bea
[[T^m,P^q],P^r]+[[P^q,P^r],T^m]+[[P^r,T^m],P^q]=0.
\eea
Plugging M charge from (\ref{meremass}) leads to equation:
\bea\label{jacobim}
f(m+q)=-\frac{(lm+m+q)}{(lm-q)}f(q).
\eea
Now, we try to interpret $f(q+2)$ in terms of $f(q)$. In one hand we have:
 \bea\begin{split}
&f(q+2)=\frac{l+1+(q+1)}{q+1-l}f(q+1)=\frac{l+1+(q+1)}{q+1-l}\frac{l+1+q}{q-l}f(q)
\end{split}\eea
and in the other hand we have:
\bea
f(q+2)=\frac{2l+2+q}{q-2l}f(q).
\eea
So, we are left with the following constraint:
\bea\label{constraint}
\frac{l+1+(q+1)}{q+1-l}\frac{l+1+q}{q-l}=\frac{2l+2+q}{q-2l},
\eea
which trivially can't be held by different values of $q$. So, any extension as of (\ref{meremass}) for infinite extended algebra is impossible. The only possibility is to have an infinite set of generators. Inspired by Schr\"odinger-Virasoro algebra we can write such charge as: 
 \bea\begin{split}\label{generalm}
&[T^m,P^q]=(lm-q)P^{m+q} \cr&[T^m,M^n]=-nM^{m+n},\cr&[P^p,P^q]=(p-q)M^{p+q}.
\end{split}\eea\\ 

\underline{\bf Case 2b}: M Charge for finite l-Galilei algebra\\

We observed that for infinite extended algebra the constraint of Eq. (\ref{constraint}) can't be held and a central charges as of Eq. (\ref{meremass}) is impossible. For finite algebra however we always can save this charge from Jacobi identity constraint. For finite algebra in Eq. (\ref{jacobim}), $"m"$ only can take the values of 0 and $\pm1$. So, $f(q+1)$ can be expressed in terms of $f(q)$ and since $q$ ranges from $-l$ to $l$, in agreement with holographic studies\cite{Galajinsky2} we find that the central charge needs to be expressed as:
\bea\label{regularm}
[P^{-l+i},P^r]=M(-1)^i\frac{i!(2l-i)!}{(2l)!}\delta_{-l+i+r,0}.
\eea
For integer $l$ we have a choice of $i=l$ anr $r=0$. Through this choice we find that for integer $l$ we need to set $M=0$. Thereby such extension is only possible for half-integer $l$. For $l=\frac{1}{2}$ or Schr\"odinger algebra this charge is nothing but the well-known mass central charge.


\section{Concluding Remarks}

Despite conformal symmetries which are recognized as symmetries to describe critical phenomena at equilibrium, the non-relativistic ones have been suggested to describe either dynamics of critical phenomena in some regimes or strongly anisotropic systems (see for example \cite{{Henkel1994},{Henkelbook}} and references therein). In this regard the dynamical index $z=\frac{1}{l}$ expresses either "How space is different from time" or "How two directions are anisotropic in space". Many models have been suggested for Schr\"odinger algebra (\cite{Henkel1994}) which is identified by $l=\frac{1}{2}$. Two-point correlators identified by $l=4$ symmetry have been suggested to find application in describing a certain multicritical point called "Lifshitz point" in \cite{Henkelprl}. As well application of this symmetry has been suggested for ANNNI model in \cite{Pleiming}. For discussion on $l=6$ see \cite{Henkelbook}.

In our work we tried to go one step ahead to know l-Galilei algebras. The detailed of our discussion can be summarized as:
\begin{enumerate}
\item
The T charge trivially exists for any element of the class. 
\item
A form of central extension as of B charge only holds for $l=1$.
\item
A regular central extension as of M charge is impossible for infinite-extended algebras. An extended form of mass charge used in Schr\"odinger-Virasoro algebra however is possible for all elements of the class. 
\item
Though a regular extension for mass charge is impossible for infinite-extension of any element of the class, it is however possible for finite algebras of half integer $l$.
\end{enumerate}

We investigated possibilities for central extension of all elements of l-Galilei class in $1+1$. The case of $1+1$ is important since CFT, Schr\"odinger-Virasoro algebra and CGA have found solution there. So, there is more chance that we know other elements of the class in this dimension.

\section*{Acknowledgments}
I would like to thank M. Henkel and S. Rouhani for fruitful comments on earlier draft and S. Moghimi-Araghi for discussions.




\end{document}